\documentstyle[12pt,aasms]{article}
\tightenlines

\lefthead{Martini, Tiede, \& Frogel}
\righthead{NGC 2477}

\begin{document}

\begin{center}
Accepted for publicaton in {\it The Astronomical Journal}
\end{center}

\vskip 24pt

\title{The Giant Branches of Open and Globular Clusters in the Infrared
as Metallicity Indicators: A Comparison with Theory}

\author{Glenn P. Tiede, Paul Martini, \& Jay A. Frogel}

\vskip 24pt

\affil{Department of Astronomy, The Ohio State
University, 174 W. 18th Avenue, Columbus, Ohio  43210 \\
tiede, martini, frogel@payne.mps.ohio-state.edu}

\clearpage

\centerline
{\bf Abstract}

We apply the giant branch slope--[Fe/H] relation derived by Kuchinski
{\it et al.} [AJ, 109, 1131 (1995)] to a sample of open clusters.
We find that the slope of the giant branch in
$K$ vs. ($J-K$) color-magnitude diagrams correlates with [Fe/H] for
open clusters as it does for
metal-rich globular clusters but that the open cluster data are
systematically shifted to less negative values of giant branch slope,
at constant [Fe/H].  We use isochrone models
to examine the theoretical basis for this relationship and find that 
for a given value of [Fe/H], the slope of the relationship remains constant
with decreasing population age but the relation shifts to less negative
values of giant branch slope with decreasing age.  Both of these theoretical
predictions agree with the trends found in the data.  Finally, we derive 
new coefficients for the giant branch slope--[Fe/H] relation for specific
members of 3 populations, metal-rich globular clusters, bulge stars and
open clusters.

\clearpage

\section{Introduction}

Relatively little work on galactic open clusters has been carried out in the 
near-infrared (1.0 -- 2.2 $\mu$m, hereafter IR), even though this wavelength
region is well suited for the study of objects in the plane of our galaxy where 
the effects of reddening can be substantial. 
Furthermore, the evolved and brightest stars in these clusters emit most of
their energy in this region of the electromagnetic spectrum, making the IR a
natural place to study them. In this paper we present an empirical
relation for the metallicity of open clusters based upon their IR giant
branch (hereafter GB) slope ($\Delta (J-K)/\Delta K$). This method is
an extension of the work by Kuchinski {\it et al.} (hereafter, KFTP) for
metal-rich globular clusters. 

\cite{kuc95} demonstrated both empirically
and theoretically that the slope of the upper giant branch in a
$K$ vs. $(J-K)$ color-magnitude diagram (CMD) is sensitive to the
metallicity of the population.  They investigated
this correlation for metal-rich
globular clusters and derived a linear relation between [Fe/H] and
GB slope.  This relationship was later refined by \cite{knf95} to be
\begin{equation}
{\rm [Fe/H]} = -2.98(\pm0.70) - 23.84(\pm6.83)\times ({\rm GB~slope}).
\end{equation}
This equation reproduces independently determined globular cluster
[Fe/H] values to $\pm0.25$ dex and so is comparable in quality to other
photometric
metallicity indicators.  \cite{tie95} applied this relationship to 8
fields along the galactic minor axis in the bulge.  
They found that Equation 1 reproduced independently determined metallicities
for the bulge fields to $\pm 0.29$ dex.

In this paper, we extend the [Fe/H] -- GB slope correlation to the 
giant branch of galactic open clusters.  In section 2, we describe 
how we selected our sample of open clusters and the technique for measuring
the slope of the upper giant branches.  In section 3, we 
investigate the effects of age on the correlation and derive preliminary new
analytic relations for the studied populations.  Section 4 contains our
concluding remarks and indications for future work with open clusters.

\section{Sample Selection}

By examining the [Fe/H] - GB slope correlation in the context of open cluster
giant branches, we attempt to extend the correlation from metal-rich globular
cluster and bulge populations to 
the thin disk population.  To date very little IR photometry of open
clusters exists in the literature.  Only two papers, \cite{hou92} and 
\cite{fro88}, present $J$ and $K$ photometry complete enough for the
purposes of this work.  Table 1 lists the open clusters we analyzed
along with their metallicities, reddenings, ages, distance moduli, and 
galactic coordinates.

We performed all of our analysis in a manner similar to that of 
\cite{kuc95}.  Using the reddenings and distance moduli from the respective
papers, all of the photometry was reddening-corrected and $M_K$ was
calculated.  We included only giants on the upper giant branch which have
absolute $K$ magnitudes in the range $ -2 \leq M_K \leq -6.5 $.  This
restriction excludes both clump giants and bright AGB
stars.  In addition, we excluded any stars designated as non-members or known
variables.  Finally, we included in our sample only those clusters which
contained $\geq 8$ giants that met all of these criteria; a total of 4 open
clusters.  The least-square fits to the giant stars in each of these 4 
clusters are shown in Figure 1 along with the stars 
on each giant branch.  The number of stars in each fit after a $2\sigma$
rejection, the derived slopes, and
the uncertainties in each slope are given in Table 2.  The uncertainty in the
GB slope of NGC 2204 is larger than the other 3 clusters due to larger
uncertainty in the photometry of stars at the dim end of the giant branch (see
Houdashelt {\it et al.} 1992).  Although the number
of stars on each giant branch is small compared with the globular cluster
or bulge studies, the values of the giant branch slopes derived from the
fits are comparable to the values from the other studies.

\section{Discussion}

Using the derived GB slopes from Table 2 and the relation given in
Equation 1, we calculated the [Fe/H] for each of the four open clusters
and compared the derived values with those from the literature.  Table 3
lists the values from the literature and the values calculated from
Equation 1.  The literature values are from various sources (see
Houdashelt {\it et al.} 1992, Table 1) and were determined using various
photometric techniques.  Errors given for the literature values are the
standard deviation in the spread in the values from the various sources.
Although the correlation of [Fe/H] and GB slope holds qualitatively for open
clusters (i.e. steeper slopes mean lower [Fe/H]), the [Fe/H] of the open
clusters is underestimated by Equation 1.  Figure 2a is a plot of 
the literature values of [Fe/H]
versus GB slope for globular clusters (solid circles), bulge fields (stars),
and open clusters (open circles).  The solid line is Equation 1.  From this
figure, it is apparent that the slope of the relationship
is similar for the three populations, but that there is a systematic
shift to more positive GB slope at constant [Fe/H] for the open clusters,
and to a lesser extent for all but one of the bulge fields.\footnote{The
one bulge field which falls to the $left$ of the
globular cluster relation is the minor-axis $-12^\circ$ field which is
likely to be composed of more halo/thick disk stars than bulge stars.}

Though the number of data points from each population is statistically small,
this change in absolute positioning is suggestive of a second
effect.  \cite{kuc95} examined the possibility of an age effect among the
globular clusters and concluded that for ages typical of globular clusters,
$10 \leq t$(Gyr)$ \leq 16$, any change due to the age spread would be much
smaller than changes due to expected metallicity variations and ultimately
too small to
detect.  However, open clusters have ages typically much less than 10 Gyr
so in the context of this work we consider ages spanning the range,
$1 \leq t$(Gyr)$ \leq 16$ when making comparison with theoretical models.
The upper limit is chosen to be in accordance with \cite{kuc95} and the lower
limit is chosen as the youngest age for which the GB slope -- [Fe/H]
correlation is likely to hold.  For clusters younger than $\sim 1$ Gyr
the upper giant branch does not terminate in a helium flash and tends to curve
toward hotter temperatures at higher luminosities.  Therefore, the upper
giant branch is not even approximately linear and the curvature will
overwhelm any metallicity effect.

To examine the relationship between upper giant branch slope and age, we
construct $M_K$ vs. $(J-K)$ CMDs using the latest edition of the Yale Isochrones
(hereafter referred to as the Y96 isochrones) from \cite{dem96}.  We construct
the CMDs for ages $= {1,2,3,5,10,16}$ Gyr and [Fe/H] $= {0.0,-0.3,-0.7,-1.3}$.
The Y96 isochrones do not contain $JK$ magnitudes; therefore it was necessary to
derive relationships between $J$ and $K$ and theoretical parameters.  We
calculated $(J-K)$ from effective temperature ($T_e$) using the relationship:
\begin{equation}
(J-K) = 259.44 - 137.72\log(T_e) + 18.31(\log(T_e))^2~~~~~(1\sigma_{fit} = 0.02)
\end{equation}
which we derived from the model atmospheres presented in Table 3 of 
\cite{coh78}.  These models are only defined in the range 
$3.58 \leq \log(T_e) \leq 3.72$ therefore we had to extrapolate Equation 2
for the coolest stars in the isochrones ($\log(T_e) \sim 3.52$).
This extrapolation was only necessary for isochrones with [Fe/H] $\geq -0.3$
and ages greater than 5 Gyr.  Even for these isochrones only stars
within $\lesssim 1.0$ magnitude of the giant branch tip had effective 
temperatures cool enough to necessitate the extrapolation.
We calculated bolometric $K$ corrections ($BC_K$) and hence absolute $K$
magnitude ($M_K$) from total luminosity and effective temperature using
the relationship:
\begin{equation}
M_K = -21.89 - 2.5\log(L/L_\odot) + 6.70\log(T_e)~~~~~(1\sigma_{fit} = 0.03)
\end{equation}
which we derived from the metal-rich globular cluster data presented in
Table 28 of \cite{fro83}.   From these data, we found a linear relation between
$BC_K$ and $T_e$ and combined this relation with the standard transformation
from total luminosity to bolometric magnitude to arrive at Equation 3.
The range of $T_e$ in the metal-rich globular cluster data spanned from
$3.5 \leq T_e \leq 3.7$ so Equation 3 did not require any extrapolation
when applied to the isochrone giant branches.
Once we used these relationships to calculate $M_K$ and $(J-K)$ for the
isochrones, we constructed CMDs and fit lines to the upper giant branches,
in the same absolute $K$ magnitude range as we did for the data.
The resulting giant branch slopes are plotted versus [Fe/H] in Figure 3.

Figure 3 is a plot of [Fe/H] versus GB slope for our constructed
theoretical $M_K$ vs. $(J-K)$ CMDs.  Each line is for a particular age (see
the figure caption).  Since the $J$ and $K$ values were calculated from
simple and incomplete (in parameter coverage) models, we do not expect,
nor do we find, good absolute agreement with the data in Figure 2.  What we
are interested in is the $relative$ change in GB slope at a constant [Fe/H]
as a function of population age.  The two 
left-most lines are for 16 Gyr and 10 Gyr populations.  Since all globular 
clusters are likely to fall in this range of ages, little change in GB slope
as a function of age (at any metallicity) is likely to be apparent.  This
agrees with the conclusion of \cite{kuc95}, where they find that age is not
likely to add any scatter to their relationship.  This conclusion is not true
for open clusters however.  Old open clusters have ages that place them between
the 10 Gyr line and the 1 Gyr line (the right most).  In this regime 
isochrones predict that the
contribution of age to determining the GB slope is comparable to the
contribution from [Fe/H].

This age effect is a possible explanation of the trend we found in Figure 2.
Assuming that the metal-rich globular clusters are older than the mean
age of the bulge stars, which are older that the open clusters, then the
systematic shift to more positive GB slope at constant [Fe/H] of the
bulge field data and the open cluster data are in the same sense as
that expected for their age differences.  This systematic shift would also
explain why Equation 1 systematically underestimates the [Fe/H] of the
open clusters -- the equation requires an age term.  We cannot, however,
derive an empirical relation between these three parameters -- age, 
[Fe/H], and GB slope -- because there are too little data presently 
available.  Additionally, a theoretical relation cannot be derived because
the current isochrones cannot even reproduce one of the empirical
relations precisely.  For example, if the globular cluster points from Figure 2
are placed on the isochrone models in Figure 3, the globular clusters
fall between the 3 Gyr line and the 1 Gyr line.  Further, the open clusters
span an age range of 5.3 Gyr (Table 1), yet the observed spread in GB slope
shift is only a fraction of that predicted by the isochrones.

Finally, we examine Equation 1 taking into account these new data.
Figure 2b displays the same data points from Figure 2a with their associated
$1\sigma$ errors.  The error estimates for the globular cluster
data and the bulge field data are discussed in \cite{kuc95} and 
\cite{tie95} respectively.  The errors for the open clusters are from
Tables 2 and 3 and were discussed previously.  The lines are error weighted
least squares fits to each population; globular clusters (solid line), bulge
fields (dotted line), and open clusters (dashed line).  The coefficients 
for these lines are given in Table 4 which also contains the slope of
the [Fe/H]--GB slope relation for each of the theoretical isochrones.
The error associated with each of these slopes is
due to the nonlinearity of the theoretical upper giant branches.  Note that
the average slope from the isochrones statistically agrees with the open
cluster and bulge slopes, both of which are statistically identical to
each other.  The globular cluster fit, though steeper than the other
relations, is heavily weighted by the lowest point.  This lowest point is
NGC 6712 which due to its rather low [Fe/H] and its combination of
galactic position--velocity, may not be part of the metal-rich disk
globular cluster system (\cite{cud88}). Excluding this point, the slope of
the globular cluster 
relation becomes statistically identical to those of the other empirical
relations and to the theoretical ones.  This concordance suggests that
the slope of the original calibration, Equation 1, is likely too large.

\section{Conclusions}

In this work we have examined the [Fe/H] - GB slope correlation, derived by
\cite{kuc95} for metal-rich globular clusters, in the context of 
open cluster giant branches.  We found that though the correlation is
in the same sense for open clusters as for globular clusters, for the open
clusters there is a systematic shift to more positive GB slopes at constant
metallicity.  This shift is also apparent to a lesser degree for the
minor-axis bulge fields studied by \cite{tie95}.

To attempt to understand the source of this systematic shift to more
positive GB slopes for the open clusters, we constructed a set of
$M_K$ vs. $(J-K)$ CMDs based on the Y96 isochrones.  We find that the
shift is in the same sense as that predicted for stellar populations of
younger ages.  Further, the Y96 isochrones predict that the slope of the
[Fe/H] - GB slope relation will remain constant for population ages 
spanning $1 \leq t(Gyr) \leq 16$.   Assuming the the age of the open
clusters is less than the age of the bulge stars and that the age
of the bulge stars is less than the age of the metal-rich globular
clusters, both of these predictions are found in the data.  Table 4
lists the isochrone and empirical relation slopes.  After the one possible
discrepant point from the globular clusters is excluded, the empirical
and theoretical slopes are all statistically identical.  Additionally, the
younger populations are found shifted to less negative GB slopes relative
to the older populations.  Though age is likely a third parameter,
we cannot derive a general relation between age, [Fe/H], and
GB slope because there are too little data presently available to allow 
an empirical derivation and the uncertainties in the theoretical models
are still to large.

The comparison of the [Fe/H] - GB slope correlation for the three
empirical populations and the theoretical isochrones demonstrates that
the correlation can be extended to populations other than metal-rich globular
clusters.  Both the theoretical models and the empirical data suggest
that the slope of the relation is $\sim -15.5$, less negative than the value
found by \cite{kuc95}.  The specific relation is dependent on age.  The
sensitivity to age is small for old populations like globular clusters
but becomes increasingly more sensitive at younger ages.  This explains
why \cite{kuc95} found no significant age effect while we find a
systematic offset for the younger open clusters.  Table 4 provides the
coefficients of least-squares fits to the various populations which are
likely more appropriate than Equation 1 for populations other than
metal-rich globular clusters.

\acknowledgements
The authors thank Sukyoung Yi and Pierre Demarque for providing us with
the latest version of the Yale Isochrones.  We also thank Leslie Kuchinski,
Marc Pinsonneault, and Andy Gould for helpful comments and discussion.  J.A.F.'s
research is supported by NSF Grant No. AST92-18281.

\clearpage

\begin{center}
\begin{tabular}{lcccccc}
\multicolumn{7}{c}{{\bf TABLE 1}}\\[12pt]
\multicolumn{7}{c}{{\bf Open Cluster Parameters}}\\[12pt]
\hline
\hline
\multicolumn{1}{c}{Cluster} &
\multicolumn{1}{c}{[Fe/H]} &
\multicolumn{1}{c}{$E(B-V)$} &
\multicolumn{1}{c}{Age} &
\multicolumn{1}{c}{$(m-M)_0$} &
\multicolumn{1}{c}{$\ell$} &
\multicolumn{1}{c}{$b$} \\
\hline
Melotte 66\tablenotemark{a}  & -0.51 & 0.14 & 6.3 & 13.0 & 259.6 & -14.3 \\
NGC 2204\tablenotemark{a}    & -0.38 & 0.08 & 2.8 & 13.1 & 226.0 & -16.1 \\
NGC 2477\tablenotemark{a}    & -0.02 & 0.22 & 1.0 & 10.5 & 253.6 &  -5.8 \\
NGC 7789\tablenotemark{b}    & -0.25 & 0.31 & 1.6 & 11.3 & 115.5 &  -5.4 \\
\hline
\end{tabular}
\end{center}

\noindent
$^{a}$ Houdashelt {\it et al.} (1992), Table 1

\noindent
$^{b}$ Frogel \& Elias (1988)

\clearpage

\begin{center}
\begin{tabular}{lccc}
\multicolumn{4}{c}{{\bf TABLE 2}}\\[12pt]
\multicolumn{4}{c}{{\bf Derived Cluster Parameters}}\\[12pt]
\hline
\hline
\multicolumn{1}{c}{Cluster} &
\multicolumn{1}{c}{N} &
\multicolumn{1}{c}{GB slope\tablenotemark{a}} &
\multicolumn{1}{c}{$\sigma$(slope)} \\
\hline
Melotte 66  &   9 & -0.077 & 0.014 \\
NGC 2204    &  10 & -0.096 & 0.024 \\
NGC 2477    &   8 & -0.111 & 0.006 \\
NGC 7789    &  10 & -0.099 & 0.004 \\
\hline
\end{tabular}
\end{center}

\noindent
$^{a}$ GB slope $\equiv \Delta (J-K)/\Delta K$

\clearpage

\begin{center}
\begin{tabular}{lcccc}
\multicolumn{5}{c}{{\bf TABLE 3}}\\[12pt]
\multicolumn{5}{c}{{\bf Comparison of [Fe/H] Values.}}\\[12pt]
\hline
\hline
\multicolumn{1}{c}{Cluster} &
\multicolumn{2}{c}{Literature} &
\multicolumn{2}{c}{Derived (Eq. 1)} \\
	& [Fe/H] & $\sigma$([Fe/H]) & [Fe/H] & $\sigma$([Fe/H]) \\
\hline
Melotte 66\tablenotemark{a}  & -0.51 & 0.15 & -1.14 & 0.33  \\
NGC 2204\tablenotemark{a}    & -0.38 & 0.20 & -0.69 & 0.57  \\
NGC 2477\tablenotemark{a}    & -0.02 & 0.05 & -0.33 & 0.14  \\
NGC 7789\tablenotemark{b}    & -0.25 & 0.20 & -0.62 & 0.10  \\
\hline
\end{tabular}
\end{center}

\noindent
$^{a}$ Houdashelt {\it et al.} (1992), Table 1

\noindent
$^{b}$ Twarog \& Tyson (1985)

\noindent
$^{c}$ [Fe/H] values have been compiled from various sources (see
Houdashelt {\it et al.} 1992, Table~1) and are based on {\it uvby},
{\it UBV}, or DDO systems.  Errors are the standard deviation in values from
the various sources.

\clearpage 

\begin{center}
\begin{tabular}{lcccc}
\multicolumn{5}{c}{{\bf TABLE 4}}\\[12pt]
\multicolumn{5}{c}{{\bf Slope of the [Fe/H] -- GB slope relation}}\\[12pt]
\hline
\hline
\multicolumn{1}{c}{Population} &
\multicolumn{1}{c}{slope} &
\multicolumn{1}{c}{$\sigma$(slope)} &
\multicolumn{1}{c}{intercept} &
\multicolumn{1}{c}{$\sigma$(intercept)} \\
\hline
Open Clusters        & -14.243 & 1.963 & -1.639 & 0.211 \\
Bulge Fields         & -13.613 & 5.118 & -1.692 & 0.500 \\
Globular Clusters    & -21.959 & 5.920 & -2.777 & 0.612 \\
GCs w/o lower point  & -18.844 & 6.406 & -2.442 & 0.669 \\
                     &         &       &        &       \\
Isochrones --  1 Gyr & -18.728 & 0.370 & \nodata & \nodata \\
Isochrones --  2 Gyr & -14.539 & 1.212 & \nodata & \nodata \\
Isochrones --  3 Gyr & -15.870 & 2.036 & \nodata & \nodata \\
Isochrones --  5 Gyr & -14.545 & 1.247 & \nodata & \nodata \\
Isochrones -- 10 Gyr & -15.855 & 1.041 & \nodata & \nodata \\
Isochrones -- 16 Gyr & -15.287 & 0.932 & \nodata & \nodata \\
\hline
\end{tabular}
\end{center}

\clearpage

\begin{figure}
\plotfiddle{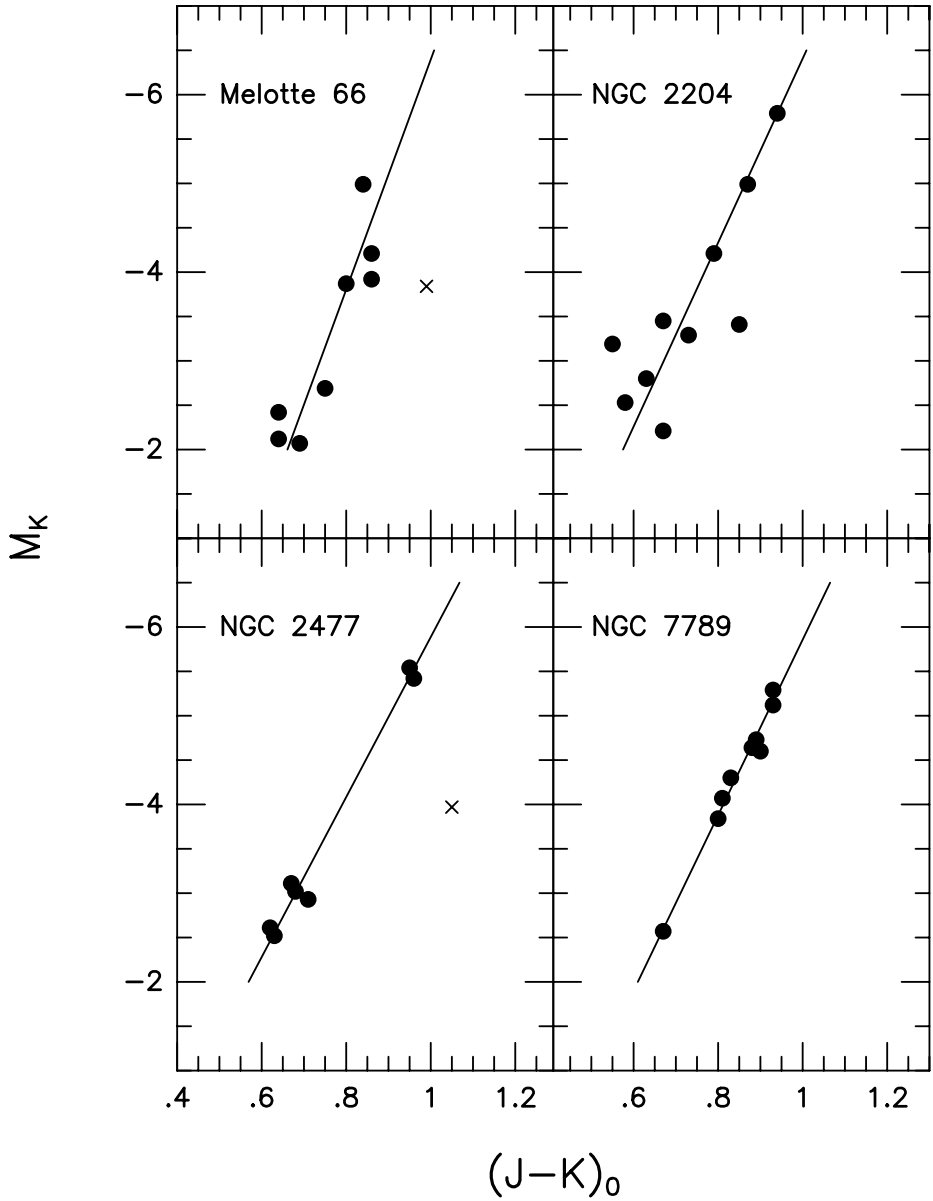}{5.0truein}{0}{70}{70}{-240}{-30}
\caption{CMD of the giant branch for each cluster studied.  The lines are
the least-squares fits to each giant branch.  Photometric errors are
comparable to the size of the points with the exception of the dimmest stars
in the NGC 2204 plot which have larger uncertainty.  Stars denoted by X's
were $2\sigma$--rejected and not included in the fits.}
\end{figure}

\begin{figure}
\plotfiddle{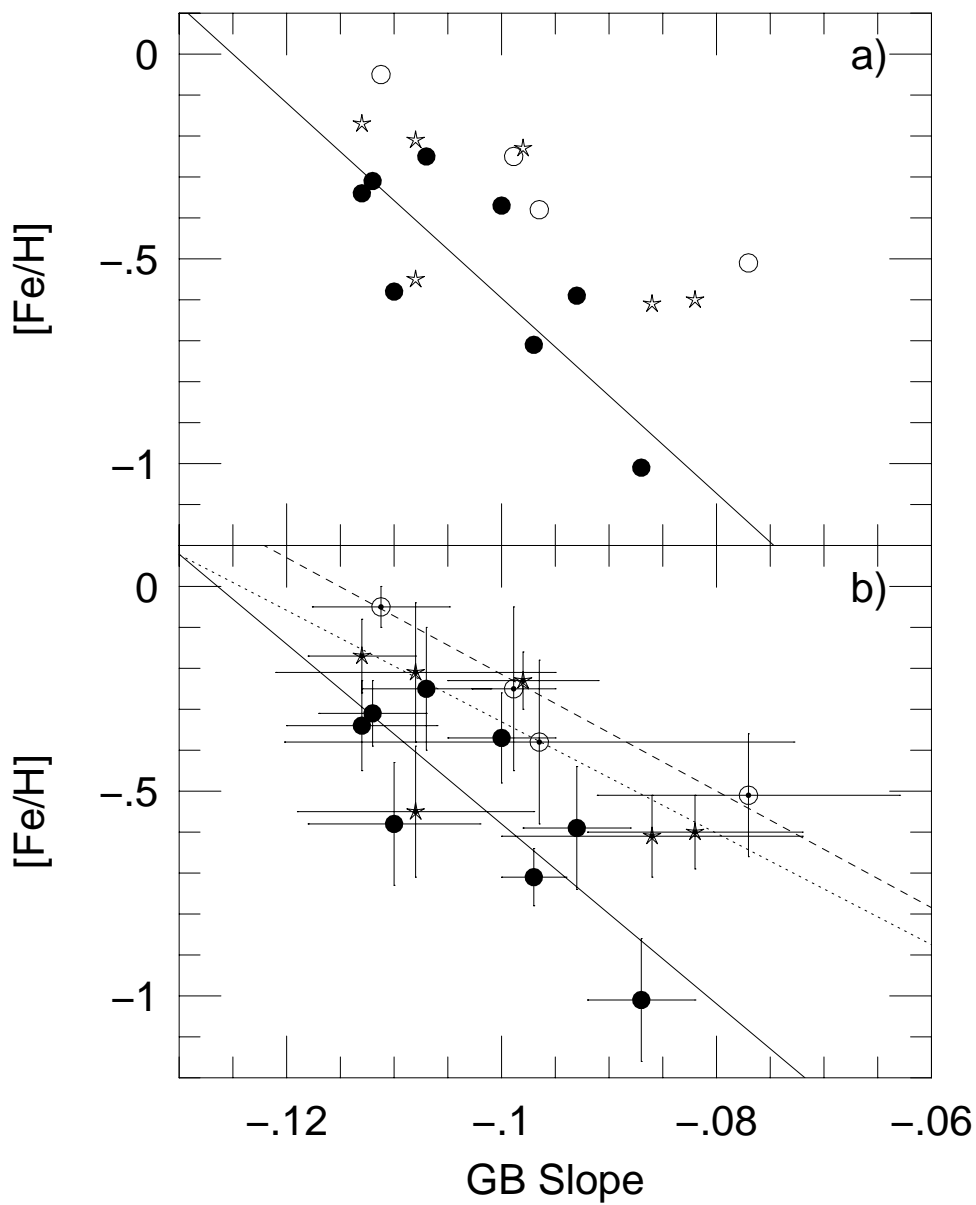}{5.0truein}{0}{70}{70}{-240}{-40}
\caption{Plots of [Fe/H] vs. giant branch slope for globular clusters (solid
circles), bulge fields (stars), and open clusters (open circles).  a) The 
solid line is the relation from KFTP derived for globular clusters (Equation 1).
b) Same data as in panel a) but with errors from the literature.  The lines are
error-weighted least squares fits for each population; globular clusters
(solid line), bulge stars (dotted line), open clusters (dashed line).
Coefficients of these fits are given in Table 4 and are discussed in the text.}
\end{figure}

\begin{figure}
\plotfiddle{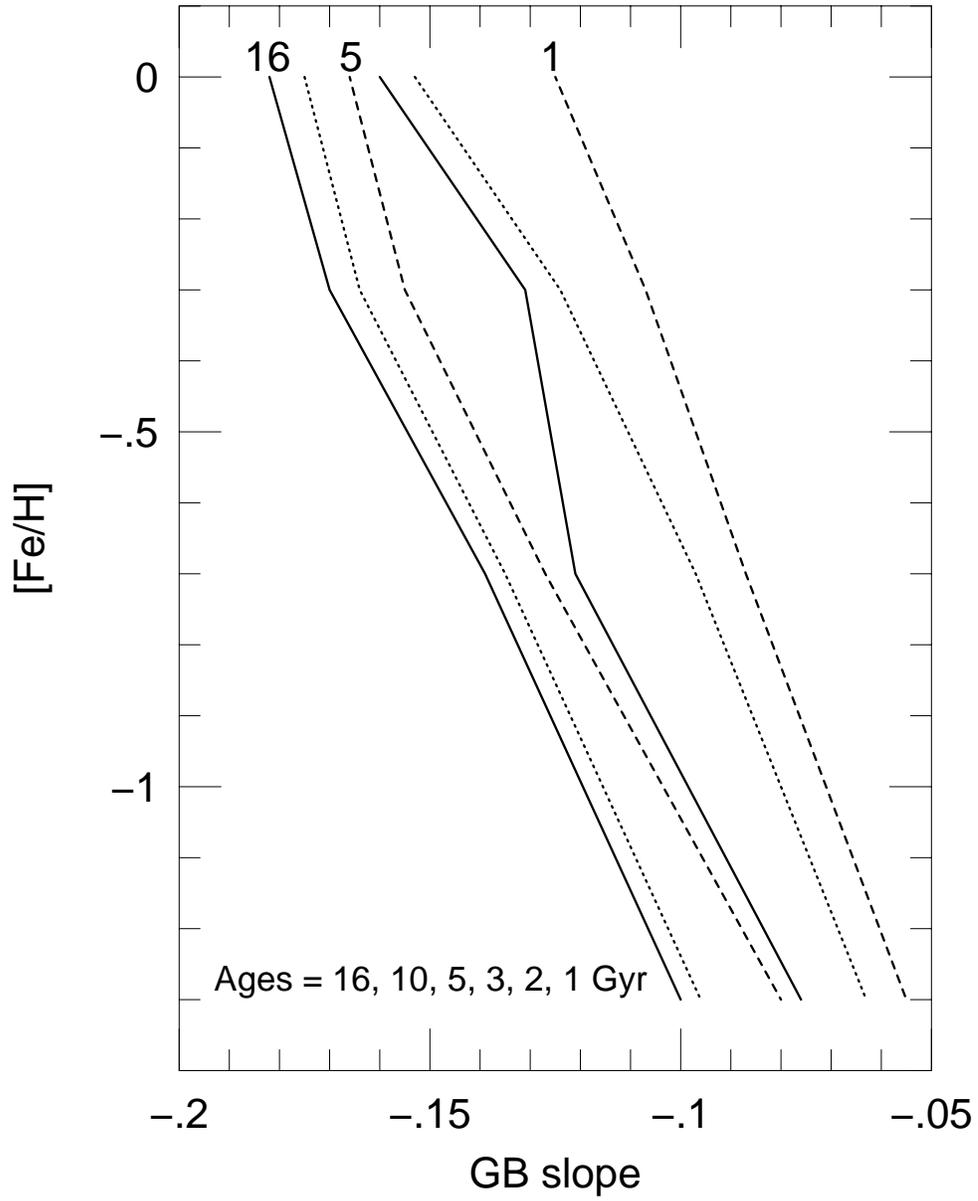}{5.0truein}{0}{70}{70}{-240}{-30}
\caption{Plot of [Fe/H] vs. giant branch slope for isochrone models at various
ages.  Ages, left to right are 16, 10, 5, 3, 2, 1 Gyr.}
\end{figure}


\begin{thebibliography}{}

\bibitem[Cohen {\it et al.} (1978)]{coh78} Cohen, J. G., Frogel, J. A., \&
Persson, S. E. 1978, ApJ, 222, 165
\bibitem[Cudworth 1988]{cud88} Cudworth, K. M. 1988, AJ, 96, 105
\bibitem[Demarque {\it et al.} (1997)]{dem96} Demarque, P., Chaboyer, B.,
Guenther, D., Pinsonneault, M., Pinsonneault, L., and Yi, S. 1997,
in preparation
\bibitem[Frogel \& Elias (1988)] {fro88} Frogel, J. A. \& Elias, J. H. 1988,
\apj, 324, 823
\bibitem[Frogel {\it et al.} (1983)]{fro83} Frogel, J. A., Persson, S. E., \&
Cohen, J. G. 1983, ApJS, 53, 713
\bibitem[Houdashelt {\it et al.} (1992)]{hou92} Houdashelt, M. L.,
Frogel, J. A., \& Cohen, J. G. 1992, \aj, 103, 163
\bibitem[KFTP]{kuc95} Kuchinski, L. E., Frogel, J. A.,
Terndrup, D. M., \& Persson, S. E. 1995, \aj, 109, 1131  (KFTP)
\bibitem[Kuchinski \& Frogel (1995)]{knf95} Kuchinski, L. E. \& Frogel, J. A.
 1995, \aj, 110, 2844
\bibitem[Tiede {\it et al.} (1995)]{tie95} Tiede, G. P., Frogel, J. A.,
\& Terndrup, D. M. 1995, AJ, 110, 2788
\bibitem[Twarog \& Tyson (1985)]{twa85} Twarog, B. A. \& Tyson, N. 1985, AJ,
90, 1247

\end{thebibliography}
\end{document}